\documentclass[conference]{IEEEtran}

\IEEEoverridecommandlockouts

\usepackage{cite}

\usepackage{amsmath}
\usepackage{enumitem}
\usepackage[vlined,ruled]{algorithm2e}
\usepackage{centernot}
\usepackage{graphicx}
\usepackage{hyperref}
\usepackage{amsfonts}
\usepackage{amssymb}

\usepackage{xargs}                      
\usepackage[pdftex,dvipsnames]{xcolor} 
\usepackage[colorinlistoftodos,prependcaption]{todonotes}
\newcommandx{\me}[2][1=]{\todo[linecolor=blue,backgroundcolor=blue!25,bordercolor=blue,#1]{#2}}
\newcommandx{\comment}[2][1=]{\todo[linecolor=green,backgroundcolor=green!25,bordercolor=green,#1]{#2}}

\usepackage{acronym}
\usepackage{multirow}
\usepackage{tikz}
\usetikzlibrary{positioning}
\usetikzlibrary{arrows}

\usepackage[normalem]{ulem}

\newif\iflong

\newtheorem{definition}{Definition}

\acrodef{OR}[OR]{Onion Routing}
 \acrodef{PPT}[PPT]{probabilistic polynomial time}

\def\PIND{P\mathit{-IND}}
\def\IIND{I\mathit{-IND}}
\def\IAIND{I\! \setminus \! |I| \mathit{-IND}} 
\def\MIND{M\mathit{-IND}} 
\def\MIIND{M \! \setminus \! I \mathit{-IND}} 
\def\RMIND{R\mathit{-IND}} 
\def\RMIIND{R\! \setminus\! I \mathit{-IND}}

\def\PINDLong{Proximity Tracing Indistinguishability}

\def\IINDLong{Infection Indistinguishability}
\def\IAINDLong{Infected Indistinguishability} 
\def\MINDLong{Meeting Indistinguishability} 
\def\MIINDLong{Healthy meeting Indistinguishability}
 
\def\RMINDLong{Remote Colocation Indistinguishability} 
\def\RMIINDLong{Remote Healthy Colocation Indistinguishability}

\makeatletter 
\newcommand\mynobreakpar{\par\nobreak\@afterheading} 
\makeatother

\hypersetup{
	colorlinks=false,
	pdfborder={0 0 0},
}

\acrodef{ACN}{Anonymous Communication Network}
\acrodef{AC}{Anonymous Communication}

\begin{document}
%
\title{Covid Notions: Towards Formal Definitions -- and Documented Understanding -- of Privacy Goals and Claimed Protection in Proximity-Tracing Services}

\author{\IEEEauthorblockN{Christiane Kuhn\IEEEauthorrefmark{1}, \hspace{1cm} Martin Beck\IEEEauthorrefmark{2}, \hspace{1cm} Thorsten Strufe\IEEEauthorrefmark{1}\IEEEauthorrefmark{3} \thanks{The authors are affiliated with, supported by, and/or members of Karlsruhe Institute of Technology, specifically the Helmholtz Centre for IT Security ``KASTEL'', Huawei, TU Dresden, and the Excellence Centre for Tactile Internet with Human-in-the-Loop ``CeTI'' (EXC 2050).}}
\IEEEauthorblockA{ \IEEEauthorrefmark{1} KIT Karlsruhe, $\{$christiane.kuhn, strufe$\}$@kit.edu \\
 \IEEEauthorrefmark{2} Huawei, martin.beck1@huawei.com \\
  \IEEEauthorrefmark{3} Centre for Tactile Internet / TU Dresden\\
  \today}
 }

\maketitle


\begin{abstract}



The recent SARS-CoV-2 pandemic gave rise to management approaches using mobile apps for contact tracing. The corresponding apps track individuals and their interactions, to facilitate alerting users of potential infections well before they become infectious themselves. 
Na\"ive implementation obviously jeopardizes the privacy of health conditions, location, activities, and social interaction of its users.
A number of protocol designs for colocation tracking have already been developed, most of which claim to function in a privacy preserving manner.
However, despite claims such as ``GDPR compliance", ``anonymity", ``pseudonymity'' or other forms of ``privacy", the authors of these designs usually neglect to precisely define what they (aim to) protect.

We make a first step towards formally defining the privacy notions of proximity tracing services, especially with regards to the health, (co-)location, and social interaction of their users. 
We also give a high-level intuition of which protection the most prominent proposals can and cannot achieve.
This initial overview indicates that all proposals include some centralized services, and none protects identity and (co-)locations of infected users perfectly from both other users and the service provider.
\end{abstract}


%
\IEEEpeerreviewmaketitle

\section{Introduction}

The current best practices to manage epidemics or pandemics in the face of highly contagious diseases is diligent contact tracing and self-isolation of potentially infected individuals so as to restrict the spread.
The extensive scale of the current SARS-CoV-2 pandemic has led to calls for automatic contact tracing.
Due to their ubiquity, the suggestions revolve around proximity detection of cell phones.

A large variety of protocols for this service have been proposed, some have passed auditing of data protection officers (the centralized mode of the Pan-European Privacy-Preserving Proximity Tracing (PEPP-PT)\footnote{https://www.pepp-pt.org/content} is one example), and some are even fully implemented and deployed (e.g., TraceTogether\footnote{https://www.gov.sg/article/help-speed-up-contact-tracing-with-tracetogether} in Singapore).
Their commonalities include the service functionality (tracing encounters and informing individuals that they are at risk, if they have been in proximity with another individual that later tested positive for the disease), their architecture (most protocols assume a server to participate in the service provision, c.f. PEPP-PT, \cite{canetti2020private , berke_assessing_2020, troncoso20decentralized}), and broadly the fact that they assume some adversaries to aim at extracting personal information from the service (sometimes also trolls who could abuse, or try to sabotage the service or its users\footnote{We focus on privacy and exclude DoS from our consideration for this report.}).

Most, if not all published proposals claim privacy, anonymity, or compliance with some data protection regulations.
However, none, to the best of our knowledge, has actually formally defined threats, trust assumptions and adversaries, or concrete protection goals.
Albeit their labeling to be ``privacy preserving'', their assumptions, measures, and achieved protection varies widely.

This greatly complicates understanding, discussion, and evaluation of the different ideas that are circulated.
The authors sometimes seem to be very focused on preventing one type of information loss, and term their solution ``private'' when this disclosure is prevented, but ignore (intentionally or not) the remaining disclosures.

We hence make a first step in the direction of systematically identifying disclosures, and defining privacy notions that proximity-tracing systems can achieve.
These notions revolve around preventing the leaking of the health conditions, whereabouts, and social interactions, or habits of the individuals who participate in the service.
We specify assumptions that fit the majority of circulated protocols first. Then we give a high-level intuition of seven privacy notions  in the flavor of indistinguishability games. Further, we go on by formally defining both the data and properties of the service, as well as the concrete privacy notions.
We also provide a very high-level, initial educated guess as to which privacy notions could, or cannot be achieved by the most prominent protocols that are currently circulated.
This initial overview indicates that all proposals include some centralized services, and none actually protects identity and (co-)location of infected users from both other users and the service
provider.

Through this endeavor we hope to help developers in improving their designs, to provide better (or: more targeted) privacy. 

\section{Background} \label{sec:background}

\subsection{Functionality}
During local infections, contact tracing traditionally is done by inquiring a patient with a positive test about the people they have met throughout the characteristic contagion period of the respective disease, and subsequently calling each of the identified individuals to ask them to self-quarantine for the incubation time.
This approach is highly privacy invasive, yet very specific: the tested individual is requested to release all contact information to the health authorities (or an organizing institution to which such services are delegated), surrenders the set of encountered individuals that still come to mind, and that the person feels comfortable to disclose. 
The institution thus gets to know these individuals and quite accurate social interactions.

Extreme situations, like the current SARS-CoV-2 pandemic, require alternative approaches. 
We observe that as incubation periods are long, it is difficult for tested individuals to remember all encounters. Further as large numbers of people are tested positive, manual interviewing and calling of encountered individuals does not sufficiently scale.

Several suggestions have been made to use location information from cell phones for this purpose.
They fundamentally comprise of three functions: sensing potential infection events, either by tracking all users' locations, or encounters between groups of users; deriving ``at-risk" (potentially infected) users upon the event of one user being tested positive; and informing the at-risk users to take further measures (e.g., self-isolate, get tested).
Note, that at least one bit of information is transferred to the alerted user, that is, the information that he or she has met one individual who has tested positive for the virus.

They commonly assume different underlying {\em roles and participating stakeholders}.  
The most basic assumption is that individuals are participating as the owners of their mobile phones (we will call their representation within the app ``users"); individuals may be untested, tested positive, or, after an encounter with a positive individual, ``at-risk''.
Testing at least in Europe is assumed to be performed by \emph{medical professionals}.
The results of positive tests are collected at public \emph{health authorities}, who also take care of informing at-risk individuals. 

\subsection{Design Space and Proposed Solutions}
Considering the design space, there is a {\em spectrum of distribution} in implementations, which come with somewhat corresponding {\em assumptions about trust} into the different entities in the system.

At one extreme there is trust into service providers and health authorities. 
Early suggestions saw carriers collect location information, knowing which cell a user is subscribed to at specific times. 
They would transmit all location information to the health authorities, who could take care of extracting encounters of individuals that have tested positive, and either directly, or through the cell phone provider contact those who are at risk.
While probably useful to trace mobility of populations, this data may be too coarse grained, as location accuracy currently is limited to the order of tens to hundreds of meters (5G will improve this to the order of centimeters, restricted to densely populated regions, though).

Location and low distance communication technologies, however, also allow the direct colocation tracing on the mobile phones of people in local proximity.
A suggestion similar to the above in regards to trust assumptions and centralization has been done in the context of Google.
Knowing the location of next to all Android and Google maps users at almost all times, Google could provide location information, derived from GPS and WiFi information (or extract encounters, when being informed about infected individuals/users) in a centralized fashion, and either inform the health authorities about individuals at-risk, or alarm them through their mobile phones directly.

Both approaches would leave some centralized entities with total knowledge of everybody's whereabouts (which they potentially already collect), and encounters, as well as information about tested and potential infections, directly tied to user identifiers or even the actual individuals.

As past experience with centralized entities has shown that total trust is not always warranted, other approaches are under discussion. 
These commonly assume the measurement and storage of encounters to be done directly on the individuals' mobile phones (most commonly by exchanging pseudonyms like ''beacons'', ''tokens'', ''Ephemeral IDs'', or other types of pseudonyms by local broadcast, and subsequently storing them on the phone).
The infected individual (upon positive testing) then commonly is assumed to surrender her local information (commonly the temporal pseudonyms from encounters she was involved in) to a service provider.
These approaches differ in the choice of centralized or decentralized information of those at risk: Centralized approaches assume the service provider to be able to map the pseudonymous colocation information from the infected user to the users or individuals at-risk (cf. the mode of PEPP-PT currently discussed for deployment in Germany), and hence to contact those directly. Decentralized approaches assume the colocation information to be published on a public bulletin board in some format that can be interpreted by all other users, to locally determine if they are at risk, or not (cf. Canetti \cite{canetti2020private}, or DP3T\cite{troncoso20decentralized}, the ``decentralized'' mode of PEPP-PT).

All of the approaches claim some kind of ''GDPR compliance'', or ''anonymity'', or to be ''privacy preserving''.
None, to the best of our knowledge, specifies precisely, what they actually mean by these terms.

\subsection{Privacy Threats}
Considering its functionality, a contact tracing system processes the (co-)location of all of its users (in some form) and the information who has been tested positive for the respective virus.
Processing these two types of information with perfect knowledge about all users clearly is privacy invasive.

Different approaches hence try to prevent certain threats: These comprise leaking {\em infection of an individual} and {\em potential infection of an individual}, for instance of all those who are alerted for being at risk.
Also leaking {\em infection} or {\em potential infection of a pseudonymous user} is a threat, as inverting a pseudonym and re-identifying  the individual may be possible, sometimes even easy, especially for long-lived pseudonyms, through different means.
Additional threats are leaking the {\em location of an individual}, even the {\em location of a pseudonymous user} (as this may allow for learning the whereabouts of another individual, tracking of a user, and as previous work has shown that pseudonymous location information can lead to very simple re-identification \cite{montjoye13unique}), or {\em encounters of two users} (as personal relationships are intimate information, and even highly distorted social graph information has been shown to be easily re-identified \cite{narayanan11link}).
Even leaking aggregate information could represent a threat, as stigmatization may happen if, for example, {\em frequencies of infections}, or {\em fractions of infected populations} become known for users with certain behavior, or in certain regions, cities, or probably even specific shops, companies, or institutions. 

\subsection{Adversary and Trust Assumptions}
With regard to the  adversaries, the common view is that the systems should protect an individual's information from the {\em health authorities} (or their delegated service providers), {\em other users}, or {\em third parties} who abuse the system for attacks on the privacy of its users, for instance by placing devices (Bluetooth antennas, video cameras, corrupted smart phones) in certain locations.
Some approaches consider attacks on the system itself, like, for instance {\em pollution} (illegitimate infection information, claiming a positive test outcome), or {\em replay/DoS} (re-broadcasting previously received tokens so as to increase the probability of this user to be alerted for being at risk). 
{\em We acknowledge the importance of attacks on the system, but limit our scope on attacks on the privacy of users within this report.}

The approaches have entirely different {\em trust assumptions}:
For the centralized PEPP-PT it is perfectly fine for the service to learn all encounters of individuals with those who are infected, but the information of who has been infected is hidden from all users. 
Even those who are alerted that they are considered to be at risk do not learn which or how many infected individuals they have met. 
Canetti, Trachtenberg, and Aria \cite{canetti2020private} (from here on {\em Canetti}) take the opposite stance and allow the alerted users to learn who of their contacts has been tested positive\footnote{By way of storing the information which token has been broadcast during encounters with which other individuals. Also storing the location for each broadcast token would even allow people who are alerted to recover the location of each individual that tests positive at the time of encounters.} but hide location traces and colocations from the service.
The same holds for the current available proposal of DP3T\cite{troncoso20decentralized}. There the server does not learn any colocation information itself, but publishes information that allows to recreate the entire list of pseudonyms of an infected user. This essentially again allows other users who stored information about the time and place of received tokens to perfectly retrace all colocations with each encountered infected user.

\subsection{Missing Formal Privacy Definitions}
{\em No formal privacy definitions and security analyses of the suggested approaches have been reported so far}, to the best of our knowledge.
Cho, Ippolito, and Yu \cite{cho2020contact} have performed an informal analysis of the TraceTogether App that is in use in Singapore, and suggest some improvements to prevent specific attacks.
Vaudenay\cite{vaudenay2020analysis} has reported some specific vulnerabilities of a version of the DP3T protocol.
Neither of the two, nor any of the protocol descriptions that we could retrieve has made an attempt to define the potential risk and formally specify the actual security goals, to make the systems accessible to formal analysis.
 
In brief, we identify a dire necessity to formalize the setting, trust assumptions, and accurate privacy notions in order to allow for a systematic understanding and discussion of the different proposals that are circulated, and their flavor of privacy which they may, or may not achieve.

\section{Application Scenario and Assumptions}
We will describe some overall terminology, the application setting, the adversaries, and threats in the following section.

Humans get infected and potentially transmit diseases.
We consider the humans to be {\em individuals}, with names ('Alice', and 'Bob').
In possession of a cell phone they can become {\em users} with corresponding pseudonyms, some of which can be (easily) linked back to the individual (the pseudonym hence {\em inverted} or {\em mapped}) - like for instance contact information (the mobile number, or the app identifier). 
For some pseudonyms this may not be simple (short term random IDs), or probably even impossible for others.
It is always easy for an individual to know its pseudonyms (even if they are automatically generated at random by their mobile app because the adversary controls the device it is easy for her to observe the pseudonyms it sends).

\subsection{Functionality}
At-risk individuals, i.e. those who have been in the proximity of another individual that has tested positive in the disease specific time span\footnote{The time span for SARS-CoV-2 could be 14 days before the positive test, or, according to the current estimate of the excretion period, five days before the onset of symptoms.}, have to be alerted of their situation. 
We assume the notification has to happen at the earliest possible time. 

We make the assumption that recovered individuals may safely delete the app from their phones, since recovering from an infection is currently believed to yield immunity\footnote{For the scope of this paper we also assume that individuals who tested positive safely self-quarantine or are in medical care, and that they disable or delete the app already during this time. Enabled apps at infected users could potentially be used for hypothesis testing (passing by somebody's home address and checking if this results in an alert), which we disregard for the moment.}.

\subsection{Roles}
We consider three entities to be involved: {\em the individuals} who install and use the app (and who may or may not become sick or at-risk), {\em the service}, who can either be the responsible health authorities, or a delegated service provider, and  potential additional {\em third parties}, who are running one or several devices to break the privacy of selected, any, or all users of the application\footnote{Note, that some approaches model the service to be distributed. PEPP-PT for instance argues that testing a user is done by professional medical personnel, who signals the information that a user has tested positive to another, different service provider; some approaches suggest distributing the service provider itself, to allow for mixing-approaches, Private Set Intersection-, or Multi-Party Computation Protocols. We believe that distributing the service may improve privacy, but we argue that the achieved privacy can be defined independent of this implementation detail.}.

The effectiveness of a proximity tracing system depends directly on its adoption, and we hence assume open groups without access limitations, so it is easy for anybody to participate\footnote{Even though it is out of scope for this work, we want to note that this does not necessarily entail zero-cost identities and hence unbounded Sybil attacks, as linking, for instance, installation identifiers to mobile devices, SIM cards, or phone numbers is possible, and sufficiently secure at least against opportunistic attacks.}.

\subsection{Data}
The service needs to consider colocation, which translates to two individuals being at the same location at the same moment.
We hence have to model locations and time.

We do not model {\em location} in terms of the coordinates, or traces of location where users have been.
Several analyses have shown that anonymizing location data, especially when sequences of positions are known, is very difficult (and most likely impossible, if any utility is expected to remain).
Privacy-preserving processing or publishing of such data hence comes at least at a great overhead.
Conforming to most of the discussed protocols (all contestants for implementation as European solutions), we model colocation as the fact that two users have been within a defined proximity of each other, no matter where this encounter has taken place.

Regarding the {\em time}, we need to distinguish between two different intervals:
\begin{itemize}
\item Batch time: Each batch groups seemingly simultaneous events. They are so short (e.g. seconds) that an adversary cannot recognize the sequence of events in them.
\item Retention time (e.g. 2 weeks): The amount of time the proximity information is needed. 
\end{itemize}

\subsection{Data Processing and Information Loss}
We deem personal information of users, and potentially aggregates thereof, sensitive and aim to analyze to which extent it is kept private (or, otherwise: lost).

The service processes data, and both functionality and service provision require transfer of certain information.
Like any relevant privacy-preserving protocol, our service must leak at least a minimum of information, as otherwise it would have no utility:
Potentially infected ``at-risk'' users learn at least the single bit of information that they are at risk, and hence have encountered one or some individuals that later have tested positive.
In addition the server, if the protocol design includes one, will learn at least the fact that it is used, so that users exist.
We do not deem service information (the existence, effort, cost, etc. of the service provision) worth protecting.

We also limit our analysis to the information that leaks {\em due to using the service}, so that an adversary learns from using the application, or controlling the service or parts of it, or combinations thereof.

We do not analyze the secrecy of external (probably private) information, as it can easily be collected by other means (side channels like setting up cameras in public places, walking around town and taking notes of observations, digital dossier aggregation).
It is important to note that we make no assumptions nor statements about such external information, and, as argued by Dwork\,\cite{dwork_differential_2006}, that even the smallest loss of information in combination with external knowledge can constitute an absolute privacy loss\footnote{It follows that an analysis as done by Vaudenay~\cite{vaudenay2020analysis} that tries to assess absolute privacy given a mixture of side channel (external) information and protocol-leaks will easily lead to attacks. However, showing achievable, non-trivial privacy goals, without a definition of privacy, is much harder in the absolute privacy model given external knowledge of the adversary.}.
Thus, absolute privacy guarantees can of course not be derived from our analysis (and most probably not from any other).

\subsection{Adversary}

The adversary might participate in any role. 
It can be a curious user, the service provider (if the protocol design includes one), or external third parties.
Our formalization of privacy goals can also be used to analyze colluding adversaries, that corrupt multiple users, or a user \emph{and} third parties  to learn private information.
Due to the assumption of open groups, an adversarial service provider can also easily create artificial users in collusion, so merge the observations at the service provider and the controlled users.

\subsection{Indistinguishability and Anonymity Sets}
Privacy terminally means that observations either cannot be made, or are impossible for the adversary to link to specific individuals.
Given observations that the adversary can make, we follow common practice and define anonymity sets:
Anonymity sets group users that are indistinguishable in a certain protocol for the adversary under a specific privacy notion. 
Information that leaks from a privacy-preserving protocol and that allows to distinguish two users, automatically puts them into separate anonymity sets (or eliminates their anonymity). If the anonymity sets thereby get so small that all users in the set have a certain property in common, then this is property is of course learned by the adversary.
This separation utilizes all information the selected adversary learns. Starting with public information (e.g., which region a user belongs to in the work of Berke et al.~\cite{berke_assessing_2020}), up to information gathered by attacking the system within the  adversarial capabilities. 

\section{Overview}

We model the privacy notions for proximity-tracing services as indistinguishability games.
An adversary submits two alternative scenarios to a challenger who chooses to execute one of them at random.
The adversary subsequently is tasked to guess which one the challenger chose, based on her observations and protocol outputs.

\begin{figure}[thb]
\begin{center}
\resizebox{0.35\textwidth}{!}{%
\begin{tikzpicture}[font=\sffamily]
    \node (PO) at (0,0) {$\PIND$};
     \node (POI) [below  =0.5cm of PO] {};
    \node (MO) [ left =1.5cm of POI] {$\MIND$};
    \node (IO)[right  = 1.5cm of POI] {$\IIND$};
    \node (MOI) [below left =0.5cm of MO] {$\MIIND$};
    \node (GO) [below right  =0.5cm of MO] {$\RMIND$};
    \node (IOI)[below = 0.4cm of IO] {$\IAIND$};
     \node(GOI)  [below right =0.4cm of MOI]{$\RMIIND$};

    \draw [semithick,->] (PO) -- (MO);
    \draw [semithick,->] (PO) -- (IO);
    \draw [semithick,->] (IO) -- (IOI);
    \draw [semithick,->] (MO) -- (MOI);
    \draw [semithick,->] (MO) -- (GO);
      \draw [semithick,->] (GO) -- (GOI);    
      \draw [semithick,->] (MOI) -- (GOI);

\end{tikzpicture}}
\caption{Hierarchy of defined notions} \label{HierarchyEarly}
\end{center}
\end{figure}
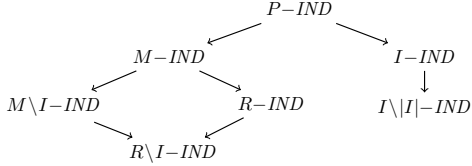

We give a brief intuition of the notions we are defining below and an overview over their relations in Figure~\ref{HierarchyEarly}.

\subsection{Intuition of Proximity-Tracing Privacy Notions}

Modeling the threats as defined above, in the context of the protocols that have been proposed and published so far, we derive the following seven relevant privacy notions:
\begin{enumerate}
	\item {\bf \PINDLong\ ($\PIND$)}:
	Other than the service functionality (at-risk individuals learn that they are at risk) and technical limitations, no entity learns anything that it does not already know by trivial observation. Technical limitations include e.g. that the service learns whether users participate at all.
	\item {\bf \IINDLong\ ($\IIND$)}: Individuals may know who they have met, at-risk individuals are informed that they met one or several infected individuals within the contagion period; but they do not learn who was infected and how many infected individuals they have met.
	\item {\bf \IAINDLong\ ($\IAIND$)}: Individuals may learn the same as in \IINDLong ~and additionally they may learn how many infected individuals exist.
	\item {\bf \MINDLong\ ($\MIND$)}: An individual may learn in how many encounters she was involved and who is infected, but does not learn who she met, nor anything about colocation of other individuals.
	\item {\bf \MIINDLong\ ($\MIIND$)}: An individual may learn the identity of infected other individuals and their colocations, but does not learn the identity of healthy others she encountered nor anything about colocations between other healthy individuals.
	\item {\bf \RMINDLong\ ($\RMIND$)}: Nobody learns anything about encounters between two benign individuals (infected or not).
	\item {\bf \RMIINDLong\ (${\RMIIND}$)}: Information about the encounters of infected users can be learned (even if they only meet benign users), but nothing can be learned about encounters between two benign, \emph{uninfected} individuals.
\end{enumerate}

We formally define each notion below, and provide several examples for valid and invalid scenarios in Appendix~\ref{app:examples}.
\section{Model}
We model events, which the adversary can choose to define the scenarios, and a notion of time that fits for this purpose.
We borrow the general idea of game based notions combined with ideas of differential privacy from existing work on anonymous communication networks and construct the game as in~\cite{ourFramework}. 

\subsection{Setting}
\subsubsection{Events}
We use $\mathcal{U}$ as the fixed  set of individuals who use the proximity tracing app, and further need to model two events:
\begin{description}
\item [\textbf{Colocation of individuals} (Meeting)] $m(A,B)$ meaning that individual $A$ was in proximity of individual $B$. We assume this event to be symmetric. Hence,  $m(B, A)$ describes the same event.
\item [\textbf{Infection of an individual}] $i(A)$ meaning that $A$ was tested positive. As we assume such users deinstall the application (the user self-isolates and is resistant to the infection afterwards), the same infection event is not allowed to occur multiple times.\footnote{As a reinfaction might be possible, we treat this event as an reinstallation of the application, which generates a new User ID.}
\end{description}

Events that follow from these, like creating pseudonyms as user identifiers, and notifying other users to be tested or self-quarantine (e.g. $B$ above), will follow from the  app protocol and are not needed for defining the scenarios (they will be, of course, important for analyzing protocols regarding the notions, later on).

\subsubsection{Time}
We need a sense of time. Therefore we assume events to be clustered into batches $\underline{r}=(r_1,\dots, r_l)$ with $r_i\in\{m(\cdot,\cdot), i(\cdot), \diamond\}$, where $\diamond$ is the empty event, i.e. no event taking place. 
All events of a batch are assumed to happen simultaneously (or at least with so little time difference that the adversary cannot observe their order). 
Every new batch contains all events of this defined very short time span\footnote{The precise time span depends on what is considered to be indistinguishable for the assumed adversary. We imagine this being in the order of a second.}. The next batch then contains the events of the following time interval of the same length. Multiple identical colocation events (e.g. $m(A,B)$ and $m(B,A)$)  are not allowed in the same batch as one cannot meet someone multiple times at the same point in time.
Further, the information is stored (and compared) for a chosen retention time; conforming to $t$ batches.

\subsubsection{Non-modeled attributes} If needed for the protocol analysis, we assume that attributes we did not explicitly model are equal  in both scenarios and are chosen reasonable, i.e. such that the chosen events might happen in the real world. For example the location at which a corrupted user meets someone in the batch is fixed and the location of her meeting in the next batch should be reasonably close. Similarly, if time is used by the protocol, it is equal in both scenarios for the same batch index.

\subsection{Game}
We first describe the easiest version of the game and then describe the additional steps that lead to the complete game.

$\mathcal{A}$ is the adversary, $Ch$ the challenger and $\Pi$ the analyzed protocol.
 \begin{enumerate}[leftmargin=1.25cm]
 \item $Ch$ randomly picks challenge bit $b$.
\item $\mathcal{A}$ sends a batch query, containing  two batches $\underline{r}_{0}$ and $\underline{r}_{1}$ (with potentially many events), to $Ch$. 
\item $Ch$ checks if the query is valid,  i.e. both batches differ only in information that is supposed to be protected according to the analyzed notion. 
\item If the query is valid, $Ch$ inputs the batch corresponding to $b$ to $\Pi$. 
\item $\Pi$'s output $\Pi(\underline{r}_{b})$, i.e. the observations according to the adversary model, is handed to $\mathcal{A}$. 
\item After processing the information, $\mathcal{A}$ outputs her guess $g$ for $b$.
\end{enumerate}

The analyzed notion ($X$) is achieved if no PPT algorithm $\mathcal{A}$ can win the game with non-negligible advantage: For all $\mathcal{A}$ exists negligible $\epsilon$, s.t.
\[ \bigm| \text{Pr}[0= \langle \mathcal{A} \mid Ch(\Pi, X, 0)\rangle ]  -
\text{Pr}[0= \langle \mathcal{A} \mid Ch(\Pi, X, 1)\rangle]\bigm|\leq \epsilon\]

where $\text{Pr}[g= \langle \mathcal{A} \mid Ch(\Pi, X, b)\rangle ]$ describes the probability that the adversary $\mathcal{A}$ guesses $g$ if the challenger chose $b$ and simulates $\Pi$ for the notion $X$.

\paragraph*{Additional Steps}
\begin{description}
\item[\emph{Multiple Batches}] Steps 2-5 can be repeated. We call the combination of all input batches for one challenge bit a \emph{scenario}; more precisely for $b=0$ the combination of input batches is the first scenario; for $b=1$ the second.
\item[\emph{User corruption}] Instead of Step 2, $\mathcal{A}$ can also decide to issue a corrupt query specifying a user $u$ and receive $u$'s internal state as output. This includes that the adversary gets access to all secret keys and stored information this user owns. This might change $\Pi$'s state, lead to different behavior of $\Pi$ in following queries and yield a higher advantage in guessing than before. Once corrupted users stay corrupted until the game ends.
\item[\emph{Protocol queries}] Any other input to the protocol (to change its state) and model active capabilities of the adversary. 
\end{description}

\subsection{Notions}
Our notions define, what the adversary is not allowed to learn.
The scenarios, which the adversary has to distinguish, hence have to be identical except for exactly the piece of information that  the corresponding notion requires to be protected.
The challenger checks for this in its check for validity, in step 3) of the game.
To easily describe the allowed differences in the input batches, we define notions as a combination of properties. These properties usually require that certain data or meta data of the real world has to be equal in both scenarios (i.e. it is not protected by the notion and can be learned by the adversary).

\subsection{Further notation}
Let $C_i$ be the set of corrupted users in the $i$-th batch, as defined by the user corruption queries above. We use $C$ as a shorthand for all $C_i$'s. Note that by definition of the corruption query $C_i\subseteq C_{i+1}$.
We use $[x,y]$ short for $\{x, x+1, \dots, y\}$ and $\underline{r}[x,y]$ as an abbreviation for all events happening in the batches $\underline{r}^x, \underline{r}^{x+1}, \dots, \underline{r}^y$.
We also use $\_$ to declare that this part of a tuple can be any value.


\section{Properties}
In this section we describe the properties, i.e. partial information about what happens in the real world or is represented in the protocol. 
This information may, or may not be desirable to hide from an adversary. 
Hiding a certain property is a characteristic of a specific protocol.
The property otherwise (most probably) can be observed by the adversary as an artifact of the protocol execution.
There certainly are numerous possible properties that theoretically could be considered. 
We restrict ourselves to those properties that we find to be most insightful, and which seem to relate to the current protocol proposals, best (Appendix~\ref{moreProperties} further elaborates the situation).

\subsection{Informal}
Informally speaking, we want a property that helps us to express that \emph{no} information leaks, other than what is unavoidable (necessary for the functionality). 
Therefore, we define the \emph{unavoidable loss} (UL) that per se has to be identical in both scenarios, as it cannot be protected. 

Further, we want to be able to analyze the protection regarding the information loss about the colocations/meetings and the infected patients. 
Therefore, we define a property ($E_m$) that allows us to easily require that no information about the colocations, but possibly any information about the infected users (except for their colocations with other users) is leaked. 
We do this by requiring that in both scenarios exactly the same infection events happen ($E_m$ are equal in both scenarios), but leaving complete freedom of choice to the adversary regarding the meeting events.

Similarly, we define another property ($E_i$) that protects only the information of infected users and fixes the colocation events.

Achieving strong notions of course is favorable, as any information leakage may be reason for concern, especially when considering auxiliary information from external sources.
Some design decisions (local broadcasting of tokens, disclosing the number of encounters to the server, or through the server to the public, etc), however, may be useful from a pragmatic perspective, yet systematically prevent protocols from achieving such strong notions against different adversaries.
We thus restrict these properties slightly by requiring for the infected protection that the number of infected users is equal in both scenarios (as this usually is published anyways) and for the colocation protection that the colocation events of infected users might leak (as some protocols leak more information about the infected users) or the colocation events of corrupted users might leak (as corrupted users learn the pseudonyms of their meeting partners and can link them to other information in some protocols). 
We summarize the intuition of our properties in Table~\ref{tab:informal}.

\begin{table}
\caption{Informal Summary of the Properties}
\label{tab:informal}
\begin{tabular}{ c  p{2cm} p{4.5cm} } \hline
 Notation & Name & Meaning  \\ \hline \hline
  $E_i$ & Equal except infected & Protects all information about infected, nothing else  \\
  $E_m$ & Equal except meetings & Protects all information about meetings, nothing else  \\ \hline
   $|U_i|$ & Number of infected& The number of infected users is not protected\\
      $|M|$ & Number of meetings& The number of meeting events happening is not protected\\ \hline
  $M_i$ & Meetings of infected& Infected peoples' meetings are not protected \\ 
    $M_c$ & Meetings of corrupted& Corrupted peoples' meetings are not protected (thus identity of meeting partner might be leaked).  \\ \hline
  $UL$ & Unavoidable loss & Information assumed to be always learned by the adversary   \\ \hline
\end{tabular}
\end{table}

\subsection{Formal Definition}
\subsubsection{Decision space}
We first define the properties where a designer potentially can decide whether this information should be protected or not.

For all following properties we assume that the adversary already sent $k-1$ batches per scenario and for these the challenger already checked that each pair was compliant to the notion analyzed. Now the next pair of batches was sent by the adversary and we define under which conditions they are valid. The current batches are thus the $k$-th batches of the game: $\underline{r}^k_0=({r_0}_1, \dots, {r_0}_l )$, $\underline{r}^k_1=({r_1}_1, \dots, {r_1}_l)$.\\

\begin{definition}[Properties]
 The current batches fulfill the  following properties if:\\
\textbf{Equal except infected $E_i$:} any event, that is not an infection event is equal in both batches\footnote{Note that it is sufficient to talk about single batches here, instead of scenarios as the properties are checked for any input batches and thus the property holds also for the complete scenarios.}. For $b\in \{0,1\}$:
\[{r_b}_j=m(\_, \_)\implies {r_1}_j={r_0}_j\text{, for all }j\in \{1, \dots, l\}\]
\textbf{Equal except meetings $E_m$:} any event, that is not a meeting event is equal in both batches. For $b\in \{0,1\}$:
\[{r_b}_j=i(\_)\implies{r_1}_j={r_0}_j\text{, for all }j\in \{1, \dots, l\}\]
\textbf{Number of infected is equal $|U_i|$:} The number of infected users is equal in both scenarios.
\[|U_0|=|U_1|  \text{, with } U_b=\{u \mid i(u) \in \underline{r}^k_b \}\]
\textbf{Number of meetings is equal $M$:} The total number of meeting events is equal in both scenarios.
\[|M_0|=|M_1|  \text{, with } M_b=\{r \mid r=m(\_, \_) \in \underline{r}^k_b \}\]
\textbf{Meetings of infected are equal $M_i$: } No one can be infected, if their meetings in the last $t$ batches differ in the two scenarios. Therefore, we define the set of users whose meetings in the two scenarios differ at some point in the last $t$ batches (i.e. at least one user is involved in the meeting of one scenario that is not in the other)  as the set $U_{\overline{m}}$. This is the set of all users $u$ that were involved in a meeting (that was event $j'$ of batch $j$) which differs in at least one user in the two scenarios.
\begin{align*}
U_{\overline{m}}:=&\{u\mid \exists j \in [k-t+1, k], \exists j': \\
&m(u_1,u_2)={\underline{r}^j_0}_{j'} \land m(u_3,u_4)={\underline{r}^j_1}_{j'} \land \\
&(u_1\neq u_3 \lor u_2 \neq u_4) \land (u \in \{u_1,u_2,u_3,u_4\})  \}
\end{align*}
We define $M_i$ to be true iff for the set of infected user $U_b$ ($U_b$ defined as in Number of infected $|U_i|$):
\[U_{\overline{m}} \cap U_b= \emptyset\text{.}\]

\textbf{Meetings of corrupted are equal $M_c$:} No one can be corrupted, if their meetings in the last $t$ batches differ in the two scenarios.
$U_{\overline{m}}$ defined as above.
We define $M_c$ to be true for the current set of corrupted users $C_k$ iff \[U_{\overline{m}} \cap C_k= \emptyset\text{.}\]

\end{definition}

%
%

\subsubsection{Unavoidable information loss}\label{ULC}
We define properties that cannot be hidden from the adversary as it belongs to the intended functionality or we consider it inherent in the technical solution that is assumed.
This is a delicate choice, as it implicitly -- and inaccurately -- accepts that some information cannot be hidden (we will comment on these decisions, below).

At-risk users always learn at least one bit due to the function of the service (that they met one or some infected users). 
Consider, for example, a user that met only one other user and then is warned. 
This trivially discloses that the individual she met is infected.
More generally, also if more users are met: Unless one limits the functionality\footnote{by e.g. allowing false-positives, false-negatives or delaying the warning} an adversary can always trivially distinguish the two scenarios, if her choice leads to notification (or lack thereof) for different corrupted users in the two cases. 
To prevent our game adversary to break our notions based on this trivial attack type, we add the property $M^{I \rightarrow C}$  and require notified, corrupted users to be equal in both scenarios. Recall that $C_i$ describes the set of corrupted users in batch $i$.
\begin{definition}
\textbf{At risk corrupted users are equal $M^{I \rightarrow C}$:}  We want that the same corrupted users  get a notification for being at risk in both scenario. Therefore, we define the set of users that both get notified and are corrupt, to be equal.
The users that get notified are those that have in the last $t$ batches been in contact with any infected user $u_i$ (whose infection is detected).
\begin{align*}
N_b:=&\{u \in \mathcal{U} \mid \exists u_i\in \mathcal{U}: i(u_i) \in \underline{r}_b^k \\
&\land( m(u, u_i) \in  {\underline{r}_b}^{[k-t+1, k]} \lor  m(u_i, u) \in  {\underline{r}_b}^{[k-t+1, k]}\} 
\end{align*}
Thus requiring the subset of the corrupted notified to be equal is:
\[N^C_0=N^C_1\text{, where }N^C_b:=N_b \cap C_k \text{.}\]\\
\end{definition}

If a corrupted user meets another corrupted user, she also knows her identity ($M^{C\rightarrow C}$).
\begin{definition}
\textbf{Corrupted meeting the same corrupted $M^{C \rightarrow C}$:} The same corrupted users have been meeting with each other in the last $t$ batches.
\[N^{C'}_0 =N^{C'}_1 \text{, where}\]
\begin{align*}
N^{C'}_b:=&\{(c_1,c_2) \mid c_1,c_2 \in C_j:  \\
& (m(c_1,c_2)\in  {\underline{r}_b}^{[k-t+1, k]}  \lor m(c_2,c_1) \in  {\underline{r}_b}^{[k-t+1, k]})\}\\
\end{align*}
\end{definition}

Further, if a corrupted user is tested positive, the adversary learns this information ($U_i^{C\rightarrow}$).
\begin{definition}
\textbf{Infection events of corrupted users $U_i^{C \rightarrow}$:} Every corrupted user is infected at the same time in both scenarios.
\[{U_i}_0^C={U_i}_1^C\]
\begin{align*}
 {U_i}_b^C:= &\{ u\in C \bigm|   i(u)\in {\underline{r}_b}^{k} \}\\
\end{align*}
\end{definition}

Additionally, if we assume broadcasting of temporal identifiers\footnote{Note that other assumptions about the technical solution lead to adapted unavoidable leakage.} any user learns how many encounters  they had. We thus define $Q_m^{C\rightarrow}$ such that the adversary cannot trivially break the notions based on this knowledge. 
\begin{definition}
\textbf{Frequency in meeting for corrupted users $Q_m^{C \rightarrow}$:} Every corrupted user had the same number of colocation events in the last $t$ batches in both scenarios. Therefore, we define the set of encounters for a user $u$ as:
\begin{align*}
M_u:=&\{j\in[1,l], i\in[k-t+1,k]\mid \\
& \exists u': m(u,u') \lor m(u',u) ={ {\underline{r}_b}_j}^i\}
\end{align*}
That the amount of encounters per user has to be the equal  for the corrupted users in the last $t$ batches is thus:
\[Q_0^C=Q_1^C\text{, where }  Q_b^C := \{ (u,n) \bigm|  u\in C_k,  n=|M_u| \}\]\\
\end{definition}


If we assume broadcasting of temporal identifiers also the fact whether multiple corrupted users met the same user at the same time, is always leaked:
\begin{definition}
\textbf{Meeting with corrupted $M_{c-c}$:} If multiple corrupted users met the same user in one scenario, they also all meet one user in the other scenario (although this user might be a different one with the same health status).  Let $U_0$ be the infected users of the first and $U_1$ of the second scenario (as in the definition of the number of infected).
For all corrupted users $c,c'\in C_k$ meeting an infected user $u\in U_0$, event $j, j' \in [1,l]$ it holds that:
\begin{align*}{r_0}_j &\in\{m(c,u), m(u,c)\} \land {r_0}_{j'} \in \{m(c',u), m(u,c')\}\\
 &\implies \exists u'\in U_1: {r_1}_j\in\{m(c,u'), m(u',c)\} \\
 & \qquad \land {r_1}_{j'} \in \{m(c',u'), m(u',c')\}  
 \end{align*}
 Similar for all corrupted users meeting a healthy user.\\
 \end{definition}
 
The combination of these above properties in \ref{ULC} (the notified corrupted users, the meetings between corrupted users, the detection of infections from corrupted user, the frequency in meetings for corrupted users, and meeting with corrupted) are trivially available to the adversary. 
We hence define the combination to be unavoidable loss: 
\begin{definition} \textbf{Unavoidable loss $UL^C$} 
\[UL^C:= M^{I\rightarrow C} \land M^{C\rightarrow C}\land {U_i}^{C\rightarrow} \land Q_m^{C\rightarrow} \land M_{c-c}\]
\end{definition}

\section{Notions}
Any combination of the above\footnote{Of course also any combination of potentially other, additional properties (like e.g. from Appendix~\ref{moreProperties}) defines a notion.} properties, either fixed or defined by the adversary, constitutes a notion in principle.
We refrain from enumerating all and instead focus on the combinations that relate to claims or seemingly achieved protection in the most prominent proposals that have been circulated, as they seem most relevant.

We illustrate all notions with a short example. Horizontal lines thereby symbolize the end of a batch, $t>2$ and user $A$ is corrupt.

\subsection{\PINDLong\ ($ \PIND$) }

The adversary only learns what she trivially already knows or has to learn,  but nothing else. This is the strongest notion, given that the assumptions on the functionality and technological limits hold. 
Therefore, the batches have to fulfill the following  property \[ \PIND: UL^C\]
Example:

\begin{center}
\begin{tabular}{ c | c}
Scenario 0 & Scenario 1\\ \hline \hline
m(A,B) & m(A,C)\\	 
m(C,E) & $\diamond$ \\
  m(A,C) &   m(A,D)\\ \hline
   m(A,C) &   m(A,E)\\ 
   i(B) & i (E) \\
   i(D) & $\diamond$ \\ \hline
\end{tabular}
\end{center}

%
%
%

\subsection{\IINDLong\ ($\IIND$)} 
The strongest notion only protecting infections. The batches only differ in who is infected and we exclude the trivial attacks as above. 
Therefore, the batches have to fulfill the following  properties \[\IIND:E_i\land UL^C\]
Example:

\begin{center}
\begin{tabular}{ c | c}
Scenario 0 & Scenario 1\\ \hline \hline
m(A,B) & m(A,B)\\	 \hline
m(A,B) & m(A,B)\\	 \hline
  m(A,C) &   m(A,C)\\ \hline
   i(B) & i (C) \\ 
      i(C) &  $\diamond$ \\\hline
\end{tabular}
\end{center}

Notice that this also protects e.g. how many users are infected, how often someone was colocated with an infected individual and how many infected individuals someone has encountered.
Thus all users, that had contact to all notified corrupted users define the anonymity set. 
In case of many corrupted users working together and immediate notification after detection of an infection, this set might still be small in reality. 

\subsection{\IAINDLong\ ($\IAIND$)}
This notion is a little bit weaker than $\IIND$ as it additionally allows the number of infected app users to be learned by the adversary. 
Therefore, the batches have to fulfill the following  properties \[\IAIND:E_i\land |U_i| \land UL^C \]
As infection numbers are usually published (albeit cumulated and after some delay), we do not consider this privacy loss as very critical.

\begin{center}
\begin{tabular}{ c | c}
Scenario 0 & Scenario 1\\ \hline \hline
m(A,B) & m(A,B)\\	 \hline
m(A,B) & m(A,B)\\	 \hline
  m(A,C) &   m(A,C)\\ \hline
   i(B) & i (C) \\  \hline
\end{tabular}
\end{center}

\subsection{\MINDLong\ ($\MIND$)} 

This is the strongest notion that only protects meetings. 
The batches can only differ in who met, but not who is infected. Therefore,  the batches have to fulfill the following  properties
\[\MIND:E_m\land UL^C\]

Example: 

\begin{center}
\begin{tabular}{ c | c}
Scenario 0 & Scenario 1\\ \hline \hline
m(A,B) & m(A,C)\\	 \hline
  m(A,D) &   m(A,D)\\ \hline
  m(C,D) & $\diamond$\\ 
   i(B) & i (B) \\ 
   i(C) & i (C) \\  \hline
\end{tabular}
\end{center}

This notion does not define protection of who is infected, but who someone met, and in consequence where someone has been.

\subsection{\MIINDLong\  ($\MIIND$)}

We now relax the above. 
This notion only protects the meetings of healthy people, the identity or location of infected individuals may be disclosed. 
As soon as someone is infected, her whole meeting history is allowed to leak. 
This means only users whose meetings are identical before can be infected (as stated earlier we assume that infected users self-isolate and have no later meetings).
Therefore,   the batches have to fulfill the following  properties
\[\MIIND: E_m\land M_i\land UL^C \]

Example:
\begin{center}
\begin{tabular}{ c | c}
Scenario 0 & Scenario 1\\ \hline \hline
m(A,B) & m(A,B)\\	 \hline
  m(A,C) &   m(A,D)\\
  m(C,D) & $\diamond$ \\\hline
   i(B) & i (B) \\  \hline
\end{tabular}
\end{center}

%

\subsection{\RMINDLong\  ($\RMIND$) } Except for meetings with corrupted users and the number of edges in the colocation graph\footnote{This notion can be easily adapted to protect the number of edges in the colocation graph by removing $|M|$.}, no information about it leaks.
Therefore,   the batches have to fulfill the following  properties
\[\RMIND : E_m \land |M| \land M_C \land UL^C\]
 Example:
\begin{center}
\begin{tabular}{ c | c}
Scenario 0 & Scenario 1\\ \hline \hline
m(A,C) & m(A,C)	 \\
  m(B,C) &   m(C,D)\\ \hline
   i(B) & i (B) \\  \hline
\end{tabular}
\end{center}

\subsection{\RMIINDLong\  (${\RMIIND}$)}  Except for meetings with corrupted users and the number of edges in the colocation graph, no information about the colocation graph between non infected users leaks. 
Therefore,   the batches have to fulfill the following  properties
\[\RMIIND : E_m  \land |M| \land M_C \land M_i \land UL^C\]
Example:
\begin{center}
\begin{tabular}{ c | c}
Scenario 0 & Scenario 1\\ \hline \hline
m(A,C) & m(A,C)	 \\
m(B,C) & m(B,C) \\
  m(E,C) &   m(C,D)\\ \hline
   i(B) & i (B) \\  \hline
\end{tabular}
\end{center}

\subsection{Hierarchy}

By definition of the notions, and the corresponding limitations in the valid adversarial choices, this hierarchy follows:

\begin{figure}[thb]
\begin{center}
\resizebox{0.45\textwidth}{!}{%
\begin{tikzpicture}[font=\sffamily]
    \node (PO) at (0,0) {$\PIND$};
     \node (POI) [below  =0.5cm of PO] {};
    \node (MO) [ left =1.5cm of POI] {$\MIND$};
    \node (IO)[right  = 1.5cm of POI] {$\IIND$};
    \node (MOI) [below left =0.5cm of MO] {$\MIIND$};
    \node (GO) [below right  =0.5cm of MO] {$\RMIND$};
    \node (IOI)[below = 0.4cm of IO] {$\IAIND$};
     \node(GOI)  [below right =0.4cm of MOI]{$\RMIIND$};

    \draw [semithick,->] (PO) -- (MO);
    \draw [semithick,->] (PO) -- (IO);
    \draw [semithick,->] (IO) -- (IOI);
    \draw [semithick,->] (MO) -- (MOI);
    \draw [semithick,->] (MO) -- (GO);
      \draw [semithick,->] (GO) -- (GOI);    
      \draw [semithick,->] (MOI) -- (GOI);

\end{tikzpicture}}
\caption{Hierarchy of defined notions} \label{HierarchyExtended}
\end{center}
\end{figure}
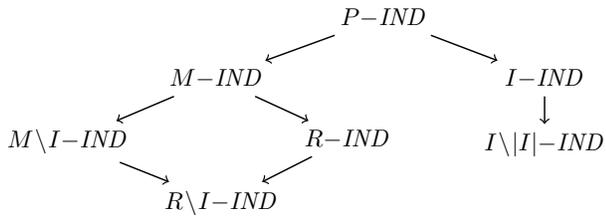

This hierarchy especially means that if $\PIND$ is achieved, any other defined notion is achieved as well; and if $\RMIIND$ and $\IAIND$ are both not achieved, none of the defined notions is achieved. Further, the notions protecting all meeting information $\MIND$ and $\MIIND$ also imply the notions protecting the respective social graph ($\RMIND$ and $\RMIIND$) and are hence stronger and better connected to each other than to the notions concerned with differing infection events ($\IIND$ and $\IAIND$). However, there might be useful weaker notions that are both implied by notions concerned with meeting information \emph{and} by notions concerned with infection information, e.g. how many meetings an infected user had earlier.

\section{Notes on existing approaches}
\begin{table*}[ht]
	\caption{An initial approximation for proposed protocols (formal proofs, and in many cases even exact protocol specifications do not exist yet)}	\label{tab:comparison}	\center
	\resizebox{1\textwidth}{!}{%
		\begin{tabular}{ c  c  c  c  c c c c c c c  c}
			\hline
			Proposal & Adversary &  $\PIND$ & $\IIND$ & $\IAIND$  & $\MIND$ & $\MIIND$ & $\RMIND$& $\RMIIND$& Location?&Pseudonyms reused?\\
			\hline\hline			
			central tracing server & client(s) &\checkmark & \checkmark &  \checkmark &  \checkmark & \checkmark  &\checkmark & \checkmark & exact & forever\\
				     & server & X &  X & X & X  & X& X & X & exact & forever \\
				     \hline
			PEPP-PT  &  client(s) & \checkmark & \checkmark & \checkmark & \checkmark& \checkmark&  \checkmark & \checkmark & no& $n^{th}$ part of day (15min)\\
				     & server & X &  X & X & X  & \checkmark &  X & \checkmark & no&$n^{th}$ part of day (15min)\\
			\hline
						Canetti &  client(s) & X& X & X & X& \checkmark& \checkmark & \checkmark & no& never\\
				     & server & X &  X & \checkmark & \checkmark  & \checkmark& \checkmark & \checkmark& no& never\\
			\hline
						MIT (PSI) &  client(s) &X &   X & X & X& \checkmark& \checkmark & \checkmark& region&  no pseudonym\\
				     & server & X &  X & X & X  & \checkmark& X & \checkmark& region&  no pseudonym\\
			\hline
						DP3T &  client(s) & X&   X & X & X& \checkmark& \checkmark & \checkmark& no& $n^{th}$ part of day\\
				     & server & X & X & \checkmark & \checkmark  & \checkmark& \checkmark & \checkmark& no& $n^{th}$ part of day\\
			\hline
	\end{tabular}}
	\label{tab:properties}
\end{table*}
Analyzing a protocol regarding a notion is only meaningful under the assumption of a certain adversary and regarding different adversaries the results for the same notion and protocol can differ. For this first short review we are concerned with the adversary model that corrupts only clients and the one that corrupts only the server.

For our educated guess, we utilize existing implications between the notions, i.e. if we expect that a strong notion is achieved this includes that all weaker, implied notions are achieved as well. Further, we use a similar logic for adversaries; if a weak adversary (e.g. one corrupting only one client) is expected to break the notion, a strictly stronger adversary (e.g. one corrupting multiple clients) is too.

We are limiting our short review of the state of the art in three ways:

\paragraph{Reusing of Pseudonyms ignored}
As long as the broadcast pseudonym is not changed, multiple meetings can trivially be linked (see Appendix~\ref{app:tracing} for a corresponding notion). This means that for this time span the user's meetings are traceable, if the adversary is able to observe the area, or at least manages to incur several encounters. 
As the information how long a pseudonym is reused is important, we include this parameter in the table. 
Other than that, we ignore this attack vector for the moment, i.e. we assumed that any batch uses new unlinkable pseudonyms, to get a feeling for the protection introduced by the other design decisions.

\paragraph{Location Information ignored}
Some protocols leverage location information to calculate the at-risk users. 
As this of course is an important private information, we include it in our table as well. 
Other than that, we however ignore this information at the moment. 
In terms of our security games  we thus assume that the locations do not differ between the scenarios, i.e. the same location is being used for the corresponding meeting of the second scenario as the one used in the first scenario. 
Also the location of one user in one batch does not change and locations of the same user in two consecutive batches are assumed to be reasonably close.

\paragraph{Educated Guess}
We have not performed in-depth analyses of the protocols in detail, and do not present any proofs. 
This is partially due to time\footnote{We believe that providing the formalization of privacy goals as notions is more pressing at this point in time, to help the various developers understand and express their goals and approaches better.}, partially due to the fact that explicit specifications sometimes are not public (we rely on a project draft for the centralized PEPP-PT) or moving targets (DP3T is under heavy development at the moment, several other approaches are published in technical reports with frequent updates).
We merely provide an educated guess of reasons, why some approaches systematically may, or may not achieve specific notions. A summary of this guess can be found in Table~\ref{tab:comparison} (the results for one and multiple clients are expected to be identical).


\subsection[title]{Centralized location tracing server\footnote{The naive solution assumed in the background (Section~\ref{sec:background}). Note, that Google and Apple recently advertised they wanted to develop their own co-location system. There currently are no specifications available but from what they hint they seemingly plan following the Canetti approach, which we discuss below, and not a centralizes location tracing server; https://www.apple.com/covid19/contacttracing/}}
\paragraph*{Clients} This approach may protect perfectly from corrupted clients as they do not have to share information with other clients, but only with the server.

\paragraph*{Server} The server learns the locations of all users at all times. Thus, it learns even more information than the colocation events that we modeled here.
\paragraph*{Notions}
$\PIND$ should hold for multiple adversarial clients as they never exchange any information with non adversarial clients and only learn unavoidable information (that they are at risk, if this is the case) from the server.
$\RMIIND$ and $\IAIND$ are broken for an adversarial server, as it learns all meeting events and infection status for all users. This implies that all other notions are broken as well. 

Further of course the exact location information is leaked and identities are known to the server.

\subsection[title]{PEPP-PT (central comparison)\footnote{We rely on an unpublished project draft.}}
\paragraph*{Client}
As clients only learn the temporary identifiers (and we ignore trivial linking due to pseudonym reusal during extended periods of time, for the moment), clients only get notified about whether they are at risk, we expect that one client cannot learn anything about the meetings or who is infected.
Multiple clients can however still learn about meeting the same user as the same pseudonym is broadcast to all.

\paragraph*{Server\footnote{The server is assumed to be trusted in this setting: The centralized PEPP-PT description argues that at-risk individuals can be alerted directly by the service, who consequently needs to know who they are. The server thus learns everything about all individuals who test positive and all their encounters, our notions hence do not match the adversary model of the developers in a very useful manner.}}
The server learns all short-term pseudonyms seen during all encounters of users who test positive, and the long-term pseudonym of the latter.
The short-term pseudonyms can be inverted and hence directly mapped to the corresponding users.
Literature shows that linking long-term pseudonyms (app IDs, cell phone numbers) to the identity of individuals is trivial, or at best represents a very light obstacle to any strategic adversary, especially with knowledge about the ego-network of the respective victim.


\paragraph*{Notions}
Corrupted clients only learn the tokens of each batch and the bit about whether they are at-risk, as the fact that multiple users learn that they are meeting the same user due to the broadcast is considered unavoidable leakage, for the moment. As the tokens cannot be linked (again: we ignore the reusage of the pseudonyms for 15 minutes in this analysis) and the bit is considered unavoidable information, $\PIND$ is expected to be achieved (and thus also any other defined notion), even against multiple clients in such a restricted analysis.

A corrupted server is expected to break $\RMIND$ and $\IAIND$. As the server notifies the at-risk users and thus learns about their meeting with infected individuals, it learns about the meetings between these users (and most likely even their identities). As the notified honest users can differ in both scenarios for $\IAIND$, this notion is also expected to break. Only the notions that allow the adversary to learn about the meetings of infected $\RMIIND$ and $\MIIND$ are expected to hold, as the server only learns the short-term pseudonyms infected users met and nothing about the meetings between non-infected users is shared with the server.

\subsection{Canetti \cite{canetti2020private}}
\paragraph*{Clients} Non-infected users only share one-time pseudonyms with other user, so we expect the meetings of them to be nearly completely private.
Infected users send all own temporal identifiers to the server.
All users receive the list of pseudonyms of infected users, as it is stored on the server.

\paragraph*{Server} The server receives all the identifiers that the infected users sent out.

\paragraph*{Notions}
$\MIND$ and $\IAIND$ are expected to break for a single adversarial user, as the user learns in which batch the match with an infected user occurs and thus the user has a good guess which the infected user was (knowing when she met him: $\IAIND$) or when she met him (knowing when the other user gets infected:$\MIND$).
We expect $\MIIND$ and $\RMIND$ to hold against that adversary, as the adversary does not learn about meetings of non-infected users, or meetings she is not involved in. As only the own keys of the infected users are published, it does not contain meeting information.

$\IIND$ is expected to break for an adversarial server, as it can observe how many batches of tokens she gets and thus learns the number of infected users. However, given the tokens are sent to the server in a protected way (via a secure anonymous communication network), the server cannot infer the identity from the tokens or any meeting information (as above). We thus expect a corrupted server (not colluding with any clients) to achieve the other notions. 

Note that the publishing of \emph{all} tokens of the infected users leaves them traceable by multiple corrupted users, if only one user has been infected in this period (if there have been several, they create an anonymity set). This should be analyzed in future work (see Appendix~\ref{app:tracing}).

\subsection{MIT \cite{berke_assessing_2020}}
\paragraph*{Clients}
Clients are assumed to have met if they are in the same region.
Clients do not directly exchange information, every communication happens over the central regional server.
Infected clients use generalization to obfuscate their location and time, before regularly publishing this information to a database.
Non-infected clients run a Private Set Intersection (PSI) or PSI-Cardinality (PSI-CA) protocol together with the database server to find out whether they have been at the same obfuscated location at the same time span as one of the infected clients. We assume that this PSI does not return duplicated set entries (no multi sets). The client learns either the obfuscated location and time for any match in the database (PSI) or the number of matches (PSI-CA).
\paragraph*{Server}
The server receives regular input from infected users, so he learns the number of infected users.
As obfuscated location-time tuples are generated in a deterministic way, infected users that meet in the same generalized location at the same time will generate the same information to be sent to the server.
The server thus learns a histogram over the number of infected people over time that meet at the same location.
Depending on the chosen PSI or PSI-CA scheme, the server also can (on its own, or together with a colluding client\footnote{He can simply take part in the open world app as a user.}), perform a bruteforce attack on the time span and obfuscated location reported by any infected user.
\paragraph*{Notions}
Note that the hashed obfuscated location-time value stored at the server and received by the clients can be efficiently resolved to the obfuscated location and time by a dictionary first preimage attack (in polynomial time in the security parameter).

$\MIND$ and $\IAIND$ are expected to break for corrupted clients as they can learn in which batch they had contact to a corrupted user (as above for Canetti). The other (not yet implied to be broken)  notions are expected to hold for this adversary as corrupted users learn no meeting information about users they did not meet or that are not infected, as they can only check for their regional information against regions infected have been in.

$\RMIND$ and $\IIND$ are expected to break for an adversarial server, as he learns the (obfuscated) location and time of infected users and can reconstruct a partly social graph and number of infected users. Further, $\IAIND$ is expected to break as the server can learn about two infected users that they have met (been at the same region at the same time) and thus can construct a scenario, where these met and another where the two infected (different users than in the first scenario) did not meet, even though in both the same number of users is infected and the same users (identities) met.
We expect $\MIIND$ to hold, as the server does not get information about the meetings of non-infected users (if the PSI is secure).

Note that the description of possibly achieved notions does not leverage the location stored in the obfuscated location-time tuple. This is due to our assumption that location of adversarial users is identical in the same batch for the moment and should be analyzed in future work.

\subsection{DP3T \cite{troncoso20decentralized}}
\paragraph*{Clients}
Clients continuously broadcast a cryptographic token via bluetooth. This token is derived from a temporary key. The key is supposed to have a validity of roughly one day. The next key is derived from the current temporal key. Each temporal key generates $n$ tokens, which are used one after another over the lifetime of the temporal key. Clients receiving such tokens are storing them together with a timestamp and signal strength. In case of an infection, the app of the infected user sends the temporal key together with a validity timespan to the server. Once a day, each client receives the server list of keys and their validity. It generates all belonging tokens and possibly finds matches with recorded tokens.
\paragraph*{Server}
The server receives a tuple consisting of a temporal key and its validity in case of an infection. Clients request the list of tuples once a day.
\paragraph*{Notions}
As above (for Canetti and MIT), $\MIND$ and $\IAIND$ are expected to be broken for a corrupted client, as she learns in which round she met an infected user (she calculates the matches on her own).  However, we expect the other (not yet implied to be broken) notions to hold, as again nothing about the non-infected meetings, or meetings that do not involve a corrupted user, can be learned.

Similar to the case with Canetti, an adversarial server learns the number of infected users (from the number of keys it gets sent) and thereby breaks $\IIND$. However, other than that the server only learns the keys of infected, but no further meeting information and thus we expect the other notions (not yet implied to be broken) to hold.

Note that the publishing of keys of the infected users allows to derive all sent tokens of this user and leaves her traceable by multiple corrupted users. This should be analyzed in future work (see Appendix~\ref{app:tracing}).


\subsection{Summary}
On this high-level overview the different approaches already show trade-offs between trust in different entities of the application and the connected privacy leakage. 

With complete trust in the server (the central location data collection, e.g. the ``Google Maps'', approach) it is relatively easy to ensure our strongest privacy goal against adversarial clients, even if several of them collude. 
Limiting the information the server learns about colocations (the PEPP-PT approach), multiple clients might infer some information about other users' meetings\footnote{The information inferred by multiple clients in the PEPP-PT approach is in our current analysis considered unavoidable leakage.}, while the server still learns enough to break nearly all our notions; the only information protected, and hence notions achieved, relate to the encounters between healthy, benign users. 
By moving the calculation of a warning to the clients (Canetti approach, DP3T) we think it is possible to achieve most of our notions against a corrupted server, at the cost of even a single client breaking some of our notions. 
Interestingly, a completely different approach (MIT), calculates the meetings and warnings at the central server, but is expected to achieve (in our current notions) the same protection from corrupted clients.
At the same time it however puts more trust and information at the server.

Slightly more general insights are that re-use of pseudonyms will likely improve the possibilities of any adversary to break notions, as throughout the reusal time the colocations can be linked, and  the corresponding user hence can be excluded from the anonymity sets of other encounters at the same time.
Another point is that we consider observed pseudonyms to be unavoidable leakage, which is only due to the fact that we assume the tokens to be transmitted via a broadcast medium (which all promising contestants for a privacy-preserving solution in Europe currently are suggesting, using 128-bit BLE tokens).
It fundamentally is possible to share them unicast in an encrypted fashion. Bystanders then cannot observe them, and each pseudonym relates to only the tuple of two users that participate in the concrete encounter.
Solutions following such an approach are expected to provide better protection against colluding clients.

Further, we have repeatedly sketched attacks against our strong notions by exploiting that a user learned the fact \emph{that} and \emph{when} she met an infected user. With the current functionality and the goal to calculate the warning at the client this seems challenging to avoid. An easier circumvention of this problem would be to allow for added delays and collect information to infer warnings for this time\footnote{The concrete analysis of this would however need an adapted notion that forces the adversary to include multiple infection events in this time span.}.



\section{Possible Extensions}
\paragraph*{Other Unavoidable Leakage}
A relaxation of the functionality or assumptions of the used techniques would lead to other unavoidable leakage. Especially that the frequency of meetings for corrupted  users and the meetings with corrupted is inherent in the underlying assumption that broadcasting techniques for some temporary identifiers are used, which the most important proposals do, but it is not inherent in the desired functionality.

\paragraph*{Different Levels of Meeting Intensity and Notifications} Our meeting event can be extended by a parameter that specifies the level of intensity or duration of a meeting. For the notions one would then again need to specify which meeting intensity each of the properties considers.

\paragraph*{More timing information for the infection} If health workers define the point of being contagious differently than $t$ time units before the detection, this information can be modeled as part of the infection event and the notions that consider the calculation of warnings can be adapted to make statements about the corresponding times.

\paragraph*{Adding Locations} Challenging the adversary to distinguish scenarios based on the location at which users are, is certainly interesting in the future for approaches that collect some form of location data. For the approaches that do not collect such location data in the beginning, such notions are trivially achieved. Modeling locations such that only reasonable inputs can be given by the adversary makes this a challenging task for future work, e.g. the locations of the same honest user in consecutive batches should be close, people that are sufficiently close should have a meeting event and different dimensions of location (GPS coordinate and height) possibly have to be considered.

\paragraph*{Epochs of reused pseudonyms}  Many protocols are based on epochs, in which one pseudonym is used. The pseudonym is only  changed for the next epoch. If all users synchronize their epoch changes (epochs are based on a global timer), it is easy to model. We can simply understand an epoch as the encapsulation of the batches that correspond to its time and extend properties as needed to consider events of this larger time span. If however all users change epochs based on their local time, the adaption of the properties is no longer straightforward and should be subject of future work.

\paragraph*{Friends, Family and People with same Routine} If one wants to model that no protection from friends and family is needed (as one might assume they share the information of being infected anyways), one can restrict the scenarios to only differ in infected users that do not meet more often than one expects people with the same routine to meet. Similarly the additional requirement that only people with the same routine differ in their infected status in the two scenarios, can be added as restriction (if one assumes that only regular met people are at risk as completely random encounters can most likely not be identified anyways \cite{troncoso20decentralized}).

\paragraph*{Assumed behavior after detection of infection} We assumed that users with a detected infection strictly self-isolate and deinstall the application as they have no more meetings. Without this assumptions further attacks are imaginable, like testing the infection status by provoking meetings with the victim, and changes on the properties would be necessary, e.g. $M_i$ would also need to ensure that meeting of an infected user \emph{after} the infect was detected are equal in both scenarios. 

\paragraph*{Sybil-Attacks} An easy attack is that an adversarial user ensures to meet only one user (by blocking later exchange of colocation data, i.e. temporal pseudonyms). If this adversarial user receives a warning, the adversary knows that the one met user is infected. If the adversary can easily create many identities, she can do this attack for many victims. As a countermeasure against this attack sybil-protection measures might be introduced into a protocol. The effect of such measures can be translated into our model by limiting the number of clients the adversary is allowed to corrupt. However, to also observe an effect on the notions, they have to adapted such that an adversary cannot pick arbitrarily many differences between the scenarios. Such a quantitative restriction of the notions can be done similar to the number of challenges in~\cite{ourFramework}.  

\paragraph*{Relaxation of Advantage definition} As done before for anonymous communication networks, in some cases the strict advantage definition is too strong and allowing to distinguish with a non-negligible, but still very small probability is acceptable. Therefore, the advantage definition can be adapted similar to the $(\epsilon, \delta)$-differential privacy definition. However, the properties can be kept the same and the same hierarchy follows.

\section{Conclusion}

The current SARS-CoV-2 pandemic has spawned a plethora of protocol proposals for contact tracing using cell phones.
The goal is to facilitate an automated system that warns people who have encountered others who have tested positive and may have been contagious during the meeting.
All of the proposals we have found claim some form of ``privacy'', ``anonymity'', or at least compliance with data protection regulations.
However, they describe a large variety of approaches, which implicitly all seem to target entirely different protection goals.
Although a few security analyses of some approaches exist, there is no systematic definition what privacy-preserving contact tracing actually means, and no formal definition of the privacy notions that correspond to the targeted protection goals.
Formal definitions do not only facilitate a systematic security analysis, they also provide means to compare different protocol proposals with respect to the protection they actually offer, and we hope that they will help design better, more privacy-preserving protocols in future.

We hence undertook a first step towards formally defining privacy notions for proximity tracing apps.
We defined seven fundamental notions in the style of indistinguishability games, with a focus on protecting the identities, health conditions, locations and social interactions of the users, and sketched their dependencies.
Applying them to some of the more prominent contestants in the current discussion, we also gave an educated guess about how these compare.
This allowed us to learn important lessons about their design decisions, and the complexity of the problem.

Our discussion indicates, that none of the current proposals can protect identity and colocation information of infected users from both other users and the service provider.
Approaches that suggest to detect potential contagion on the users' phones allow colluding adversarial clients to trace infected users. 
Those that suggest this function to be performed on the server allow the service provider to learn the set of alerted users, and leak some colocation information to the provider.
We also learned that the choice to broadcast tokens will greatly complicate protection from colluding adversaries, and that frequencies of encounters are much harder to hide, than we initially would assume.

We assumed the use of unique, short-term pseudonyms, in this work, but observe the desire to reuse pseudonyms for extended periods, to ease calculating risk-scores of colocated users, with privacy.
Our current efforts hence are to extend the analysis framework to cover reuse of pseudonyms in a meaningful manner.
The framework currently models encounters as the simple colocation of two individuals. 
Some approaches recommend to use traces of actual positions of single individuals to extract encounters by comparing such traces.
Despite the fact that the current majority opinion seems to be that such approaches will be either insecure or extremely costly, we are also aiming to integrate such data into our model.

Our initial attempt already helped to classify the approaches, and provide insight about the protected properties and trade-offs.
We would like to stress, that our discussion does not represent an in-depth analysis, yet -- some protocols are still in the process of redesign, for other protocols no public specifications exist, and that we have not proven any of the statements, so far.
A future formal analysis with our notions will help uncover more intentional and unintentional leaks. 
It can therefore provide tremendous value to the discussion and choice of protocols for deployment, and help to design  better protocols in the future.

\bibliographystyle{abbrv}
\bibliography{articles.bib}

\newpage


%

\appendix

\subsection{More Properties (Overview)} \label{moreProperties}
We systematically discuss ad-hoc properties interesting for privacy notions of proximity tracing applications in this section. We limit this version to an informal introduction of them. A formal definition similar to~\cite{ourFramework} is future work.

\subsubsection{Infections}
Two protectable properties only concern infection events:
\begin{description}
\item[Set of infected users $U_i$:] Who is infected (all or a subset of the infected users) is known or not known to the adversary. 
\item[Number of infected $|U_i|$:] How many app users are infected is known or not known to the adversary. The number of infected is usually made public in regular time intervals. However, an earlier or more fine grained information about this might be desirable to hide.
\end{description}

\subsubsection{Meetings}
Properties that only concern user meetings and no infections:
With regard to single users:
\begin{description}
\item [Complete meeting info per user $CI_u$:] 
This information contains a list of all meeting events this user was in. This includes who a user met, how often and in which order. 
\item[Meeting sets $M$:] This only includes which other user each user met, but not how often or in which order.
\item[Frequency in users $Q_u$:]
 This specifies for any user, how many users a she met.
\item [Frequency per user $Q$:] 
This specifies for any user, how often she met any other user.
\item [Frequency in meetings $Q_m$:]
 This contains for any user, the number of meeting events she had.
\end{description}

With regard to multiple users:
\begin{description}
\item[Complete meeting information $CI$:] This lists all meeting events that happened. Compared to $CI_u$ this also allows to infer the order of meetings from different users.
\item[Partitioning $P$:] 
This specifies the subset of users, in which each user  met at least one other user of the set (connected components in the meeting graphs) in a given time span.
\item[Histogram $H$:]
 This specifies how many users met how many users (e.g. 10 users met 1 user each, 5 users 2 etc. ).
\end{description}

\subsubsection{Combined}
Properties that arise form  meeting infected users:
With regard to single users:
\begin{description}
\item [Complete meeting info restricted to infected $CI_u^{\rightarrow I}$:]  This lists all meetings with infected users the considered user had. It includes which infected user she met, how often and in which order.
\item[ Meeting set restricted to infected users $M^{\rightarrow I}$:] This specifies which infected users a user met.
\item[Frequency in infected users $Q_u^{\rightarrow I}$:] This specifies how many infected users a user met.
\item [Frequency per infected user $Q^{\rightarrow I}$:] This specifies how often a user met a certain infected user, for each infected user.
\item [Frequency in meetings with infected users $Q_m^{\rightarrow I}$:] This specifies how many meeting events with infected users each user had.
\end{description}

With regard to multiple users:
\begin{description}
\item[Complete information restricted to infectes users $CI^{\rightarrow I}$:] This lists all meeting events with infected users involved.
\item[Histogram restricted to infected users. $H_m^{\rightarrow I}$:] This specifies how many users met how many infected users.
\end{description}

In short, we restrict the properties above to only take the infected users into account $X^{\rightarrow I}$ in the combined properties. 

We can also turn this around to describe which infected user met which other users and restrict the statistics on the infected users $X^{I\rightarrow}$.
With regard to single users:
\begin{description}
\item [Complete meeting info for infected $CI_u^{I \rightarrow}$] lists all events in which an infected user is involved (as $CI_u^{\rightarrow I}$).
\item[Meeting sets for infected $M^{I \rightarrow}$] specifies which users any infected user met (not the same as  $M^{\rightarrow I}$, but includes same information). 
\item[Frequency in user per infected $Q_u^{I \rightarrow}$] specifies how many users an infected user met (different information than $Q_u^{\rightarrow I}$).
\item [Frequency per user for infected$Q^{I \rightarrow}$] specifies how often an infected user met another user (different information than $Q^{\rightarrow I}$).
\item [Frequency in meetings for infected $Q_m^{I \rightarrow}$] specifies how many meeting events with users each infected user had (different information than $Q_m^{\rightarrow I}$).
\end{description}

Similarly, we can restrict this for contact of infected users under each other: $X^{I\rightarrow I}$.

\paragraph*{Note on other user sets}
Similarly, we can restrict these properties on any other subset of users. Especially, the corrupted users are useful in the unavoidable leakage: $X^{\rightarrow C}$, $X^{C\rightarrow}$ and $X^{C \rightarrow C}$.

\subsection{Tracing of users}\label{app:tracing}
A typical goal is that the colocation events of a user at different times cannot be linked to each other, as otherwise the complete movement of the user could be traced by an adversary.  To capture this protection in a notion, we need a property that ensures that in the first scenario the same user  was met twice  and in the second two different users were met. Example ($A$ corrupted):
\begin{center}
\begin{tabular}{ c | c}
Scenario 0 & Scenario 1\\ \hline \hline
m(A,B) & m(A,B)\\	 \hline
  m(A,B) &   m(A,C)\\ \hline
\end{tabular}
\end{center}

However as we do not want the adversary to win the scenarios by identifying the user \emph{only} in the first meeting, we need to ensure that this alternative is used in 50\% of the games:

\begin{center}
\begin{tabular}{ c | c}
Scenario 0 & Scenario 1\\ \hline \hline
m(A,C) & m(A,C)\\	 \hline
  m(A,C) &   m(A,B)\\ \hline
\end{tabular}
\end{center}

Formally, such a property requires small extensions to the game, but can be constructed similar to $T_S$ in~\cite{ourFramework}.
Note that depending on how close these two batches are, this idea is able to capture the leakage due to reused pseudonyms (if the batches are so close that still the same pseudonym is used, e.g. directly after each other as in the example above) or to cover the leakage due to linkable pseudonyms (if we force the adversary to compare batches that are in different pseudonym epochs).

\subsection{More examples} \label{app:examples}
As one could come up with arbitrary many examples, the following list is not entitled to be complete in any sense.
For all exampls horizontal lines symbolize the end of a batch, $t>2$ and $A, A', A''$ are corrupt users.

\subsubsection{\PINDLong\ $\PIND$ }

\textbf{Not Valid:} 
Different corrupted users are infected:
\begin{center}
\begin{tabular}{ c | c}
Scenario 0 & Scenario 1\\ \hline \hline
  i(A) & i(A') \\ \hline
\end{tabular}
\end{center}

Corrupted user has different number of meetings in both scenarios.
\begin{center}
\begin{tabular}{ c | c}
Scenario 0 & Scenario 1\\ \hline \hline
m(A,B) & m(A,C)\\ 
  m(A,C) &   $\diamond$ \\ \hline
\end{tabular}
\end{center}

Corrupted user meets different other corrupted users in both scenarios.
\begin{center}
\begin{tabular}{ c | c}
Scenario 0 & Scenario 1\\ \hline \hline
m(A,A') & m(A,A'')\\  \hline
\end{tabular}
\end{center}

Corrupted user gets a warning in one, but not in the other scenario.
\begin{center}
\begin{tabular}{ c | c}
Scenario 0 & Scenario 1\\ \hline \hline
m(A,B) & m(A,C)\\ 
i(B) &   $\diamond$ \\ \hline
\end{tabular}
\end{center}

Two corrupted users met different number of different people in the same batch.
\begin{center}
\begin{tabular}{ c | c}
Scenario 0 & Scenario 1\\ \hline \hline
m(A,B) & m(A,C)\\	 
m(A',B) & m(A',D)\\	 \hline
\end{tabular}
\end{center}

\textbf{Valid:} 

\begin{center}
\begin{tabular}{ c | c}
Scenario 0 & Scenario 1\\ \hline \hline
m(A,B) & m(A,C)\\ 
  m(B,D) &   $\diamond$ \\ \hline
  i(D) & i(B) \\ \hline
\end{tabular}
\end{center}

and any other valid input of any other notion.

\subsubsection{\IINDLong\ $\IIND$}
\textbf{Not Valid:} Different meetings happening.

\begin{center}
\begin{tabular}{ c | c}
Scenario 0 & Scenario 1\\ \hline \hline
m(B,C) & m(C,D)\\  \hline
\end{tabular}
\end{center}

 and any example not valid above.

\textbf{Valid:} How many users are infected?
\begin{center}
\begin{tabular}{ c | c}
Scenario 0 & Scenario 1\\ \hline \hline
   i(B) & i (C) \\ 
      i(C) &  $\diamond$ \\\hline
\end{tabular}
\end{center}

How often did someone meet an infected?

\begin{center}
\begin{tabular}{ c | c}
Scenario 0 & Scenario 1\\ \hline \hline
m(A,B) & m(A,B)\\ \hline
m(A,B) & m(A,B)\\  \hline
m(A,B) & m(A,B)\\  \hline
m(A,C) & m(A,C)\\ \hline
   i(B) & i (C) \\  \hline
\end{tabular}
\end{center}

How many infected did someone meet?

\begin{center}
\begin{tabular}{ c | c}
Scenario 0 & Scenario 1\\ \hline \hline
m(A,B) & m(A,B)\\  \hline
m(A,C) & m(A,C)\\ \hline
   i(B) & i (C) \\  
      i(C) & $\diamond$ \\  \hline
\end{tabular}
\end{center}

 and any valid example of $\IAIND$

\subsubsection{\IAINDLong\ $\IAIND$}
 \textbf{Not Valid:} Different number of infected:
 
 \begin{center}
\begin{tabular}{ c | c}
Scenario 0 & Scenario 1\\ \hline \hline
   i(B) & i (C) \\  \hline
      i(C) & $\diamond$ \\  \hline
\end{tabular}
\end{center}

and any example not valid before.

\textbf{Valid:}
How often did someone meet an infected?

\begin{center}
\begin{tabular}{ c | c}
Scenario 0 & Scenario 1\\ \hline \hline
m(A,B) & m(A,B)\\ \hline
m(A,B) & m(A,B)\\  \hline
m(A,B) & m(A,B)\\  \hline
m(A,C) & m(A,C)\\ \hline
   i(B) & i (C) \\  \hline
\end{tabular}
\end{center}

How many infected did someone meet?

\begin{center}
\begin{tabular}{ c | c}
Scenario 0 & Scenario 1\\ \hline \hline
m(A,B) & m(A,B)\\  \hline
m(A,C) & m(A,C)\\ \hline
   i(B) & i (C) \\  
      i(C) &  i(D)\\  \hline
\end{tabular}
\end{center}

\subsubsection{\MINDLong\ $\MIND$}
  \textbf{Not Valid:} Different infected.
 
 \begin{center}
\begin{tabular}{ c | c}
Scenario 0 & Scenario 1\\ \hline \hline
   i(B) & i (C) \\  \hline
\end{tabular}
\end{center}

and any example not valid for $\PIND$.

  \textbf{Valid:} 
How many meeting did an infected have?
\begin{center}
\begin{tabular}{ c | c}
Scenario 0 & Scenario 1\\ \hline \hline
m(B,C) & $\diamond$\\	 
m(B,D) & $\diamond$\\	 
m(B,E) & $\diamond$\\	 \hline
i(B) & i(B)\\ \hline
\end{tabular}
\end{center}

Who did an infected meet?
\begin{center}
\begin{tabular}{ c | c}
Scenario 0 & Scenario 1\\ \hline \hline
m(B,C) & m(B,D) \\	 \hline
i(B) & i(B)\\ \hline
\end{tabular}
\end{center}

and any valid example below.

\subsection{\MIINDLong\ $\MIIND$}

  \textbf{Not Valid:} Different meetings for infected.

\begin{center}
\begin{tabular}{ c | c}
Scenario 0 & Scenario 1\\ \hline \hline
m(B,C) & m(B,D) \\	 \hline
i(B) & i(B)\\ \hline
\end{tabular}
\end{center}

Different number of meetings for infected.
\begin{center}
\begin{tabular}{ c | c}
Scenario 0 & Scenario 1\\ \hline \hline
m(B,C) & $\diamond$\\	 
m(B,D) & $\diamond$\\	 
m(B,E) & $\diamond$\\	 \hline
i(B) & i(B)\\ \hline
\end{tabular}
\end{center}

\textbf{Valid:}

Who did a corrupted user meet?
\begin{center}
\begin{tabular}{ c | c}
Scenario 0 & Scenario 1\\ \hline \hline
m(A,B) & m(A,C)\\	 \hline
\end{tabular}
\end{center}

How many meetings happened (under honest users)?
\begin{center}
\begin{tabular}{ c | c}
Scenario 0 & Scenario 1\\ \hline \hline
m(B,C) & $\diamond$\\	 \hline
\end{tabular}
\end{center}

and any example valid under  $\RMIIND$

%
%

\subsection{\RMINDLong\ $\RMIND$} 
\textbf{Not valid:} Corrupted user meets different people.
\begin{center}
\begin{tabular}{ c | c}
Scenario 0 & Scenario 1\\ \hline \hline
m(A,B) & m(A,C)\\	 \hline
\end{tabular}
\end{center}

Number of meetings differs:
\begin{center}
\begin{tabular}{ c | c}
Scenario 0 & Scenario 1\\ \hline \hline
m(B,C) & $\diamond$\\	 \hline
\end{tabular}
\end{center}

and any example not valid above for $\PIND, \MIND$.

\textbf{Valid:} Who did meet with whom (under honest users)?

\begin{center}
\begin{tabular}{ c | c}
Scenario 0 & Scenario 1\\ \hline \hline
m(A,C) & m(A,C)	 \\
  m(B,C) &   m(C,D)\\ \hline
   i(B) & i (B) \\  \hline
\end{tabular}
\end{center}

and any example valid for $\RMIIND$.

\subsection{\RMIINDLong\ $\RMIIND$} 
\textbf{Not valid:} Different meetings of infected users:
\begin{center}
\begin{tabular}{ c | c}
Scenario 0 & Scenario 1\\ \hline \hline
m(B,C) & m(B,D) \\ \hline
   i(B) & i (B) \\  \hline
\end{tabular}
\end{center}

and any example not valid above for $\PIND, \MIND, \RMIND,\MIIND$.

\textbf{Valid:}
Who did meet with whom (under honest, non-infected users)?
\begin{center}
\begin{tabular}{ c | c}
Scenario 0 & Scenario 1\\ \hline \hline
m(B,C) & m(B,D)\\	 \hline
\end{tabular}
\end{center}

\end{document}